\documentclass[traditabstract]{aa}
\usepackage[utf8]{inputenc}
\usepackage{paralist}
\usepackage{amsmath}
\usepackage{microtype}
\usepackage{natbib}
\usepackage{graphicx}
\usepackage{epsfig}
\usepackage{txfonts}
\usepackage{color}
\usepackage{multirow}
\usepackage{amssymb}
\usepackage{epsfig}
\usepackage{xspace}
\bibpunct{(}{)}{;}{a}{}{,} 
\newcommand{\vela}{Vela~X$-$1\xspace}

\title{Footprints in the wind of \vela traced with MAXI.}
\author{V.\,Doroshenko\inst{1}\and A.\,Santangelo\inst{1}\and S.\,Nakahira\inst{4} \and T.\,Mihara\inst{2} \and M.\,Sugizaki\inst{2} \and M.\,Matsuoka\inst{2} \and M.\,Nakajima\inst{3} \and K.\,Makishima\inst{2,5}}	
\institute{Institut für Astronomie und Astrophysik, Sand 1, 72076 Tübingen, Germany\and
MAXI team, RIKEN, 2-1 Hirosawa, Wako, Saitama 351-0198, Japan\and
School of Dentistry at Matsudo, Nihon University, 2-870-1, Sakaecho-nishi, Matsudo, Chiba, JAPAN 271-8587\and
ISS Science Project Office, Institute of Space and Astronautical Science (ISAS), Japan Aerospace Exploration Agency (JAXA), 2-1-1 Sengen, Tsukuba, Ibaraki, JAPAN 305-8505\and
Department of Physics, The University of Tokyo 7-3-1 Hongo, Bunkyo-ku, Tokyo 113-0033
}

\begin{document}

\bibliographystyle{aa}

\abstract{The stellar wind around compact object in luminous wind-accreting
high-mass X-ray binaries is expected to be strongly ionized with the X-rays
coming from the compact object. The stellar wind of hot stars is mostly driven
by light absorption in lines of heavier elements, and X-ray photo-ionization
significantly reduces the radiative force within the so-called Str\"omgren
region, leading to wind stagnation around the compact object. In close binaries
like \vela this effect might alter the wind structure throughout the system.
Using the spectral data from the Monitor of All-sky X-ray Image (MAXI), we studied
the observed dependence of the photoelectric absorption as a function of the orbital
phase in \vela and found that it is inconsistent with expectations for a
spherically symmetric smooth wind. Taking into account previous investigations,
we developed a simple model for wind structure with a stream-like photoionization
wake region of slower and denser wind trailing the neutron star, which is responsible for
the observed absorption curve.}

\authorrunning{V. Doroshenko et al.}
\maketitle

\section{Introduction} \object{Vela X$-$1} is a persistently active high-mass X-ray binary
system consisting of a massive neutron star (1.88$M_\odot$,
\citealt{velamass}) and the B0.5Ib-type supergiant HD~77581 with a mass of
$\sim23M_\odot$ and radius of $\sim30R_\odot$ \citep{kerkwijk}. The hot primary
loses mass at a rate of $\sim10^{-6}M_\odot\,\mathrm{yr}^{-1}$ \citep{nagase86}
through a powerful stellar wind, which is accreted onto the neutron star located
$\sim53R_{\odot}$ away, and powers the X-ray pulsar. The pulsar is eclipsed by
the primary during every orbital cycle of $\sim8.964$\,d \citep{kerkwijk}. The spin
period of the neutron star, $P_s\sim283.5\,\text{s}$ \citep{rappaport_vela},
has remained almost constant since this discovery. The X-ray spectrum of \vela
is typical for accreting pulsars, and is well described by a cutoff power
law \citep{nagase86} with cyclotron resonance scattering features at $\sim25$
\citep{Makishima:1992p3220} and $\sim50-55$\,keV
\citep{kendziorra_vela, Makishima99}.
Similar to other wind-accreting systems, the X-ray flux in \vela is strongly
variable and may change by orders of magnitude within minutes
\citep{kreyken_vela,me_vela}. The variability is usually associated with the
clumpy nature of the captured radiatively driven wind
\citep{Oskinova:2007p560,fuerst}. The observed average X-ray flux implies a
luminosity of $\sim4\times10^{36}\mathrm{erg\,s}^{-1}$ (assuming a distance of
$\sim2$\,kpc, \citealt{nagase89}), which agrees with the estimate for Bondy-Hoyle
accretion from smooth spherically symmetric wind with an observed terminal
velocity of $\sim1100\,{\rm km\,s}^{-1}$ \citep{watanabe06}, and a mass-loss
rate of $\sim10^{-6}M_{\odot}{\rm\,yr}^{-1}$. The wind in \vela, however, is
\emph{not} expected to be symmetric, and the X-ray emission from the pulsar
is supposed to ionize the wind almost down to the surface of the primary, leading to
wind stagnation and formation of the photoionization wake trailing the neutron
star \citep{blondin,feldmeier96}. Similar studies were also performed for other sources
\citep{Manousakis12}. More recently, detailed calculations of
wind acceleration in the presence of an X-ray source have been carried out by
\cite{krticka12} with application to \vela for the stationary case. The authors
found that the wind velocity field is significantly affected by the feedback from
the accreting neutron star and anticipated an almost complete stagnation of the
wind within the ``photoionization-bubble'' around the neutron star.

Several studies point to the existence of such a feature in \vela.
\cite{nagase86} observed strong asymmetry between the ingress and egress of
the eclipse, accompained by increase in energy of Fe-K line and the absorption
edge, which \cite{feldmeier96} interpreted as enhanced scattering of X-rays in
the denser region, trailing the neutron star. Based on the large observed size of
the scattering region, these authors ruled out the accretion wake in favor of the
photoionization wake as the scattering site.

Based on the orbital modulation of the UV-resonance lines in several systems,
including Vela~X$-$1, \cite{vanLoon2001} concluded that a complex wind
structure with an ionized region around, and a photoionization wake behind the
neutron star is required to explain the high-resolution UV spectra of the source.

\cite{goldstein} and \cite{watanabe06} used high-resolution \emph{Chandra}
X-ray spectral data to constrain the ionization structure and the geometrical
distribution of material in the system. Intensities of the X-ray emission lines
of highly ionized H- and He-like ions were found to be compatible with the smooth
wind model \citep{castor}. However, most of the X-ray recombination lines were
found to come from between the neutron star and the primary, suggesting
an hanced density in this region, most likely due to photoionization-driven reduction
in radiative force. In fact, the velocity shifts of the highly ionized
recombination emission lines were found to be smaller in magnitude by several
hundred kilometers per second than the predictions for the smooth wind,which directly supports the presence of the photoionization wake.

Performing a detailed study of the global wind structure is
complicated, however, by the dramatic variability of both the wind,
and the X-ray feedback from the pulsar. Studying this variability
provides insight on details of wind acceleration and interaction with
the neutron star and is best addressed with high-quality dedicated
observations such as those presented for \vela by \cite{watanabe06} or
\cite{nagase86}, who performed detailed time-resolved spectral
analysis. On the other hand, to to trace the persistent wind
structure, not only long observations at different orbital phases, but
also observations at multiple orbital cycles are important, which
might not be feasible for dedicated observations. On the other hand,
because it is one of the brightest persistent X-ray sources in the
sky, \vela is an ideal target for monitor-type X-ray instruments such as
MAXI (Monitor of All-sky X-ray Image), which can complement existing
high-quality dedicated observations. In this work we used MAXI data to
study the average orbital phase dependence of the X-ray spectrum in
2-30\,keV in \vela with the aim of constraining the changes in
photoelectric absorption induced by the photo-ionization-driven
inhomogeneties in wind structure. We found that the observed changes
in the absorption column are incompatible with the
spherically symmetric smooth wind, and argue that the photoionization
wake trailing the neutron star may indeed be responsible for the
enhanced absorption.

\section{Observations and data analysis} MAXI is the first
astronomical mission operated on the International Space Station (ISS)
\citep{matsuoka09}. It was installed on the Japanese Experiment Module
— Exposed Facility on July~16, 2009, and has been designed to surpass
all operating X-ray monitors both in sensitivity and energy
resolution. Two slit-slat X-ray cameras scan the entire sky every
92\,min as the ISS follows its orbit. They operate in 0.5-12\,keV (the
Solid-state Slit Camera (SSC), \citealp{tomida11}), and 2-30\,keV (the
Gas Slit Camera (GSC), \citealp{mihara11}), which we used here. The
camera consists of twelve Xe-gas proportional counters with
slit-and-slat collimators limiting the field of view (FOV) to
$1.5^\circ\times160^\circ$, which scans the sky as the ISS orbits. The
counters employ resistive carbon-wire anodes to acquire
one-dimensional position sensitivity, and enable position
reconstruction for every source in the FOV to about 30$^\prime$. The
large detector area of 5350\,cm$^2$ allows many brighter sources to be
detected within one $\sim1$\,m scan. Typical daily exposures reach
3000\,cm$^{2}$\,s, which enables spectral studies when multiple scans
are combined. The relatively low uncertainty in the energy calibration
of about 3\%, the fair energy resolution $\sim18$\,\% (at 5.9\,keV),
and the robust background rejection via anti-coincidence with outer
veto cells additionally aid spectral capabilities. The camera
performance is summarized in \cite{sugizaki11}.

\begin{figure}[t]
	\centering
		\includegraphics[width=0.5\textwidth]{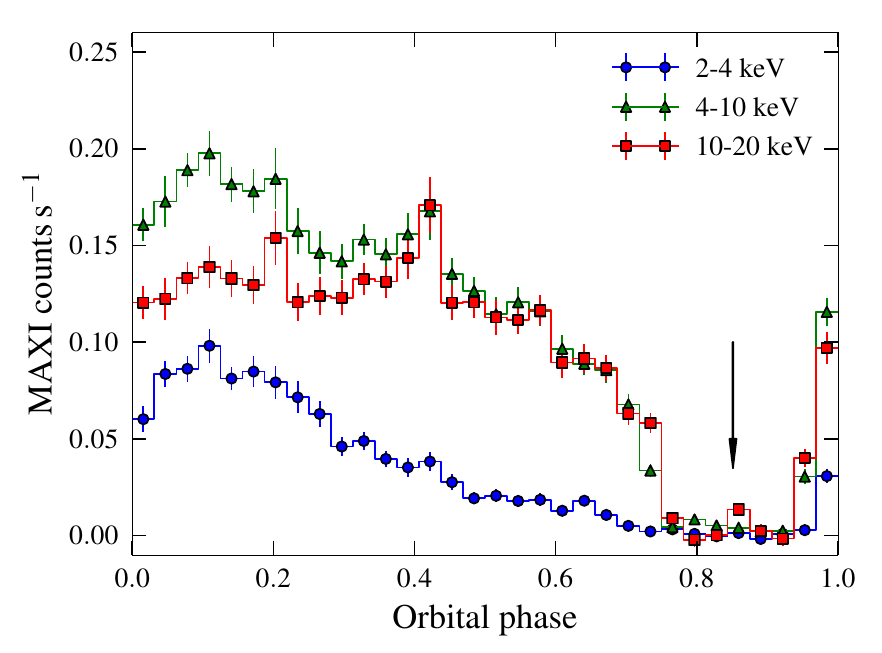}
	\caption{\vela orbital light curves in 2-4 (blue circles), 4-10 (green
triangles), and 10-20\,keV (red squares) energy ranges assuming orbital
parameters reported by \cite{kreyken_vela}, and folded from public MAXI
light curves relative to the periastron. The arrow indicates the mid-eclipse phase.}
	\label{fig:olc}
\end{figure}

The inspection of public \vela light curves provided by the MAXI
team\footnote{http://maxi.riken.jp/top} for every source detected by
the GSC in several energy ranges suggests a significant dependence of
the source spectrum on orbital phase. The comparison of the
mission-long soft and hard light-curves folded with orbital period
presented in Fig.~\ref{fig:olc} reveals that most changes occur in the
soft 2-4\,keV band, while other bands are less affected, which could
be naturally explained with the strong changes in absorption column
along the orbit. It is interesting also to note the spike in hard
X-ray flux at phase $\sim0.4$, which could potentialy provide
additional constrains on the wind structure.

To go beyond studying the hardness ratios and to verify if their change
is related to the photoelectric absorption, we performed an
orbital-phase-resolved spectral analysis of MAXI GSC data obtained
during period MJD~55141-56200 using the standard MAXI pipeline
\citep{sugizaki11,nakahira12} modified to allow arbitrary grouping of
the individual scans. The individual scan duration is just about one
minute, i.e., much shorter than the phase bin of $\sim1$ day we are
interested in, so grouping of the scans is a robust way to obtain the
orbital-phase-resolved spectra. Using the ephemeris provided by
\cite{kreyken_vela}, we calculated a list of start and end times for
each phase bin, and extracted the spectra by grouping the individual
scans accordingly. The effective exposure is uniformly distributed
along the orbit and varies between 15-20\,ks per phase bin. We rebined
all extracted spectra to contain at least 200 photons per energy bin,
and fit them in the 2-20\,keV energy range using the comptonization model
by \cite{Sunyaev:1980p2243} (\emph{CompST} in XSPEC). We kept the
temperature and optical depth the same for all spectra, and only allowed
the flux and the absorption column to vary independently. This resulted
in a statistically acceptable combined fit with $\chi^2_{red}=1.13$
($\chi=941.2$ for 831 degrees of freedom) with the best-fit values for
temperature and optical depth $kT=6.4(5)$\,keV, $\tau=13.3(7)$, and
the absorbtion column in $3.3(5)-25(4)\times10^{22}{\rm atoms\,cm}^2$
range. Estimating the continuum parameters independently does not
improve the fit significantly.

\begin{figure}[t]
	\centering
		\includegraphics[width=0.5\textwidth]{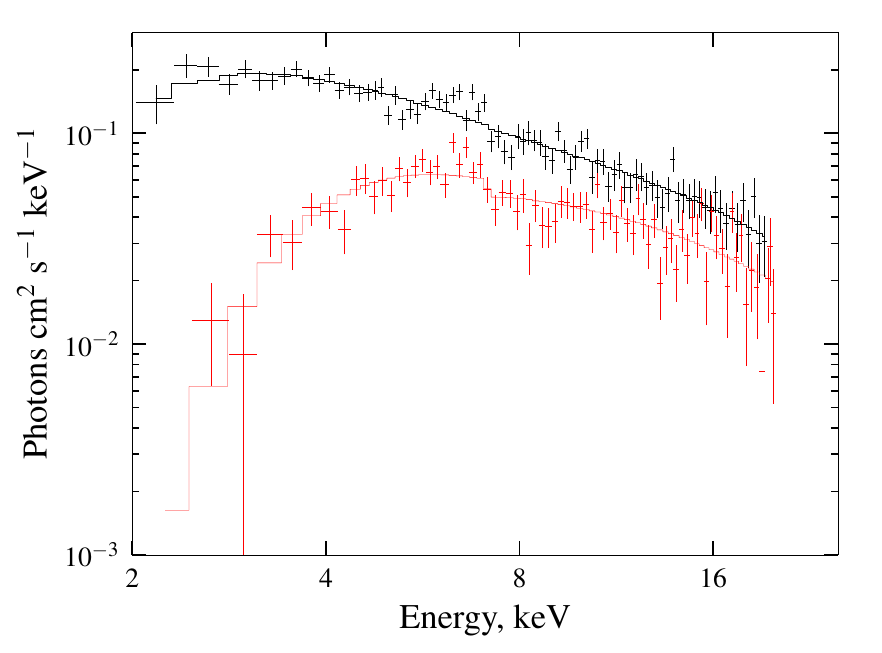}
		\includegraphics[width=0.5\textwidth]{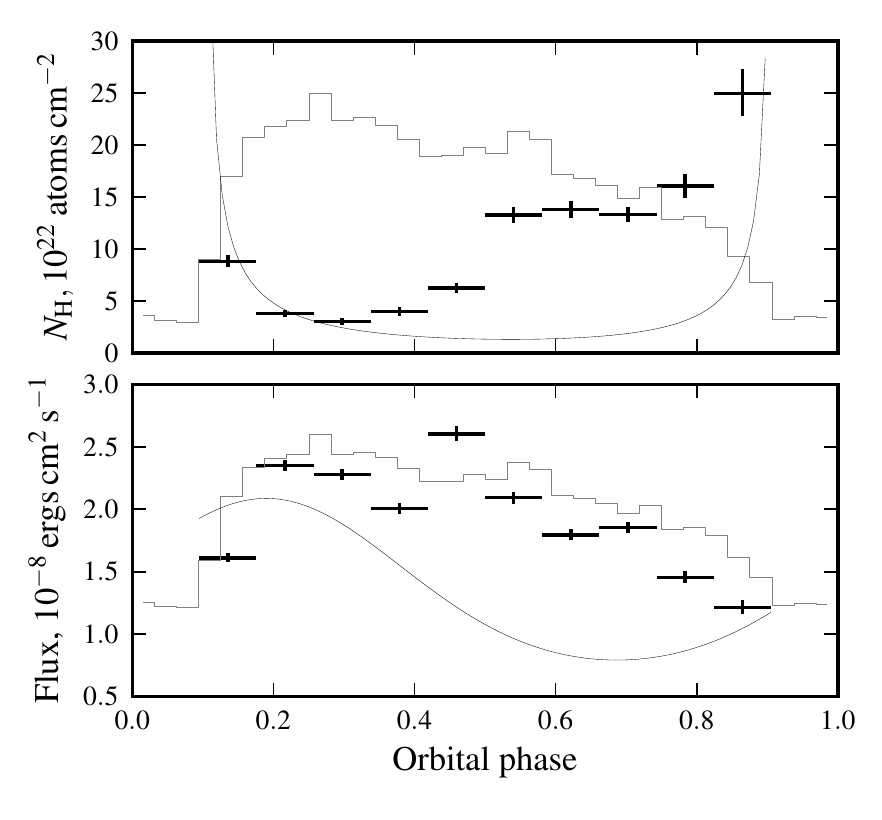}
	\caption{Representative unfolded spectra for the most (red) and least
(black) absorbed phase bins (top panel), and best-fit results for the absorption
column and flux for all spectra (bottom panel). Absorption column and flux
estimates for the smooth wind (black) and the orbital light curve (gray) are
plotted in the bottom panel. Model flux mostly changes because of change in the relative velocity of the wind and the neutron star.}
	\label{fig:spefit}
\end{figure}
Keeping the continuum parameters common for all phase bins is also justified because
the observed shape of the spectrum does not change above 10\,keV
(within available statistics). We also confirmed that the data point
distribution on a color-color plot is consistent with them being
one-dimensional, controlled only by the absorption. The representative spectra
and the orbital dependence of free parameters are presented in
Fig.~\ref{fig:spefit}. The flux plotted in the figure is unabsorbed source flux in
the 2-20\,keV range derived from the spectral fit, and uncertainties are given at
$1\sigma$ confidence level.

\section{Discussion} First of all we would like to emphasize that the observed
orbital dependence of the absorption column is asymmetric with respect to the
conjunction, and thus cannot originate in a spherically symmetric wind of the
primary. The asymmetry can only be caused by interaction of wind with the
neutron star. The observed dependence closely resembles the results of hydrodynamical
simulations carried out by \cite{blondin} that account for 
a photo-ionization wake (see for instance Figs.~8 and 11 in their work), although the
observed asymmetry is somewhat stronger in \vela. A similar orbital dependence of
the absorption column was observed in \object{IGR~J17252$-$3616} with \emph{XMM-Newton}
\citep{Manousakis11}, where the authors also attributed the enhanced absorption at
later orbital phases to a photo-ionization wake. For the same
source, \cite{Manousakis12} performed detailed hydrodynamical simulations
that confirmed their initial assumption that the peculiarly slow wind and the
photo-ionization wake trailing the neutron star are responsible for the
observed orbital dependence of the absorption column.

Indeed, comparing the observations with the results of detailed
hydrodynamical simulations is essential for understanding the effects the
accreting neutron star imposes on the wind structure. However, one needs to bear
in mind the extreme complexity of these simulations and associated limitations.
For instance, \cite{Manousakis12} estimated the mass of the neutron star based
on its impact on the simulated wind structure, which may be affected by the
assumed simplified prescription for the radiative force that drives the wind (which
is assumed to switch off within the Str\"omgren sphere). On the other hand,
\cite{krticka12} showed that the radiative force field is much more complicated
than that and has to be calculated self-consistently with the wind
structure. Performing such calculations is beyond the scope of the present
work, and taking into account that the results of a simplified approach
presented in \cite{Manousakis11} and the more detailed simulations presented in
\cite{Manousakis12} agree reasonably well, we stick to the simpler
prescription just to illustrate some important points suggested directly by the
observations.

As a first step, it is interesting to compare the observed absorption column
and flux with estimates from a standard smooth radiatively driven wind model
\citep{castor} with the radial wind velocity profile,
$$
\upsilon(r)=\upsilon_0+(\upsilon_{\infty}-\upsilon_0)\left(1-\frac{R_s}{r}\right)^\beta,
$$
where $\upsilon_0\sim10{\rm\,km\,s}^{-1}$ is the sound velocity at the surface
of the optical star with radius $R_s=30R_\odot$,
$\upsilon_{\infty}=1100{\rm\,km\,s}^{-1}$ and $\beta\sim1$ (\cite{watanabe06} and references therein).
For
the geometry of the system we assumed the orbital parameters reported by
\cite{kreyken_vela}, that the system is observed edge on (\cite{kerkwijk}
estimate $i\ge 75^\circ$), and adopted a mass-loss rate of
$2\times10^{-6}M_\odot{\rm\,yr}^{-1}$ \citep{watanabe06}. 
The density at any distance may then be calculated from the continuity
equation. This allows one also to calculate the X-ray flux assuming
the Bondi-Hoyle spherical accretion and the distance to the source of 2\,kpc
\citep{nagase86}.
The equivalent absorption column was calculated by integrating the density along
the line of sight and assuming neutral absorber cross-sections
\citep{morrison83} with solar abudances. We neglected interstellar absorption in
the direction of the source, which is estimated to be $\le5\times10^{21}{\rm
atoms\,cm}^{-2}$ \citep{Dickey90}, i.e., about an order of magnitude lower than
the lowest observed value. The results are presented in Fig.~\ref{fig:spefit}.

As previously mentioned, this simple model cannot reproduce the
observed asymmetric shape of the absorption curve along the entire
orbit. On the other hand, if the observed asymmetry is caused by some
structure trailing the neutron star, the agreement between the
observed absorption column and the model for the first three points
after the eclipse when this structure is behind the neutron star is
remarkable. The observed average flux agrees well with our simple
estimate, although its orbital dependence is not reproduced as well.
It is important to emphasize that no parameters were adjusted to
obtain the curves in Fig~\ref{fig:spefit}, which were instead
calculated using existing independent estimates for the parameters of
the wind and the binary system.

Another important point is that both the absorption column and the
flux depart from the model predictions, and that the observations
start to depart from the model \emph{before} the conjunction (phase
0.5). This means that the neutron star is not simply trailed by the
region of slower, denser wind, but is submerged within it or at least
resides at its boundary. This suggests a possible additional gradient
in wind velocity in the vicinity of the neutron star, which might have
profound effects on the amount of angular momentum captured by the
neutron star, and therefore on its spin evolution. Detailed
simulations with proper treatment of the radiative force and accurate
wind velocity calculations are required to estimate their magnitude,
however.

\begin{figure}[t]
	\centering
		\includegraphics[width=0.5\textwidth]{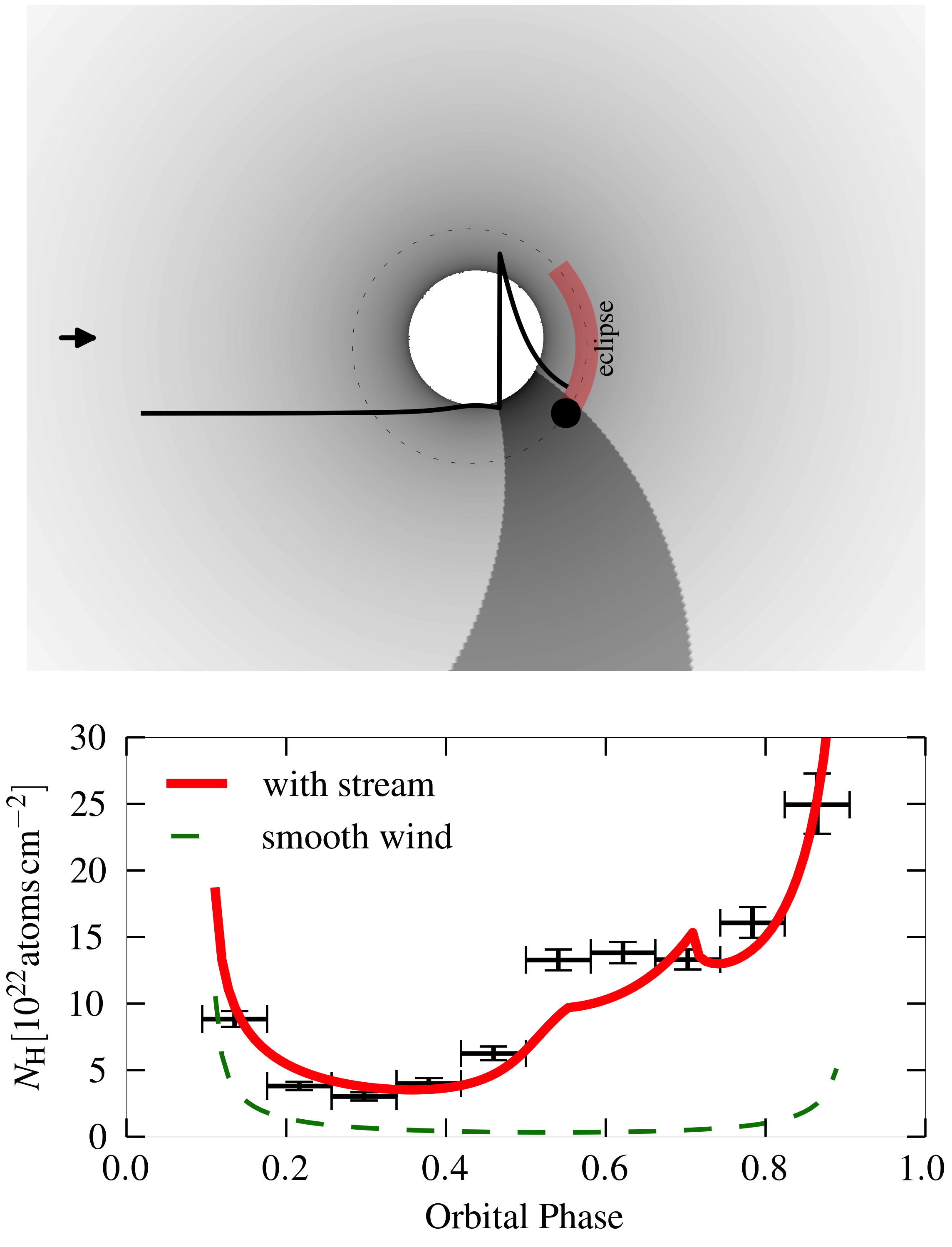}
	\caption{\emph{Top panel:} Sketch of the system with the stream trailing the neutron star as it orbits HD~77581 (top panel); the
	solid line indicates the wind density distribution along the line of sight (indicated with an arrow).
	 \emph{Bottom panel:} orbital absorption curve as observed with MAXI (crosses, uncertainties at $1\sigma$ confidence level),
	 and model curves calculated for the smooth-wind model (dashed line, see also Fig.~\ref{fig:spefit}),
	and for a model with stream trailing the neutron star (solid line). We arbitrary shifted the curve for the smooth model to enhance the plot clarity.}
	\label{fig:stream}
\end{figure}

To better illustrate these points, and to qualitatively explain the
observed changes in the absorption column along the orbit, we extended
a simple smooth-wind model by adding a bent stream of slower, denser
wind trailing the neutron star similar to the one emerging from
simulations by \cite{blondin}, \cite{feldmeier96}, and
\cite{Manousakis12}. Even these quite detailed simulations probably do
not provide a realistic picture, since the accelerating radiative
force is treated very crudely. For our ``toy'' model we just assumed
that the expected stream-like structure is described as a wind sector
with a certain opening angle ($\sim40^\circ$), the same mass-loss rate
and velocity profile as the ambient wind. The wind velocity within the
stream was, however, scaled down by a constant factor, thus increasing
the density in the stream with respect to the ambient wind by about a
factor of ten, which is calculated from the continuity equation. The
stream then bends away from the neutron star and phase lag linearly
increases with radius ($\propto0.1r$), which corresponds to rigid
rotation \citep{feldmeier96}. In reality, the wind velocity is
certainly more complex, particularly between the companions (which
implies, for instance, that the density wake will be offset from the
primary). However, the line of sight does not cross this region, and
most of the absorbing material resides in the outer wind, which has a
simpler structure. The absorption curve calculated using our ``toy''
model is qualitatively consistent with observations as presented in
Fig.~\ref{fig:stream}, and the corresponding
animation\footnote{http://astro.uni-tuebingen.de/~doroshv/stream.mov}.
The result is also quite similar to the one reported by
\cite{Manousakis11} for IGR~J17252$-$3616, which is not surprising
because they used similar assumptions. We emphasize that in our
case the model has mostly illustrative purpose, and only in conjunction
with more detailed simulations \citep[i.e.,][]{Manousakis12} it is
possible to conclude that the absrobtion in the dense stagnated wind
region in the photoionization wake is most likely responsible for the observed
absorption orbital profile in \vela as well. We did not attempt to
model the orbital flux profile here, since the accretion rate mostly
depends on the relative velocity of the wind and the neutron star, which
cannot be estimated without detailed calculations of the wind-driving
force. Taking into account the reasonable agreement of the observed
flux and estimated luminosity for unperturbed wind
(Fig.~\ref{fig:spefit}), it is clear, however, that the wind velocity in
the vicinity of the neutron star must not drascically decrease, i.e., it
most likely resides at the edge of the photoionization wake.

\section{Conclusions} We have investigated the global stellar wind structure
in \vela exploiting the continuous monitoring of the source with MAXI and its
admirable spectral capabilities. We measured the average
orbital dependence of the absorption column in the system and of the intrinsic
source X-ray flux in \vela, and found that both depart significantly from the
expectations for a smooth-wind model. We attributed these discrepancies to a photo-ionization wake, which is expected to manifest as a denser
stream-like region trailing the neutron star. We constructed a simple model to
qualitatively describe the impact of this stream on the orbital dependence of
the absorption column, and found that the neutron star must reside either
within it or at the leading edge. We found that this simple mode agrees quite
well with observations, which provide important constraints for future more
detailed wind simulations. From an observational point of view, our result also
poses a direct science case for and highlights the importance of the modern
X-ray monitoring missions like MAXI and the upcoming WFM onboard LOFT.

\begin{acknowledgements}
VD and AS thank the Deutsches Zentrums für Luft- und Raumfahrt (DLR) and
Deutsche Forschungsgemeinschaft (DFG) for financial support (grant
DLR~50~OR~0702). VD also thanks the entire MAXI team for the collaboration and hospitality
in RIKEN.
\end{acknowledgements}

\vspace{-0.3cm}\bibliography{auto_clean}	\vspace{-0.3cm}\end{document}